\documentclass[prb,twocolumn,showpacs,superscriptaddress]{revtex4}

\usepackage{graphicx}

\begin{document}

\title{Spin gaps and spin-flip energies in density-functional theory}

\author{K. Capelle}
\affiliation{ Centro de Ci\^encias Naturais e Humanas,
Universidade Federal do ABC, Santo Andr\'e, 09210-170 S\~ao Paulo,
Brazil}
\affiliation{Instituto de F\'{\i}sica de S\~ao Carlos,
Universidade de S\~ao Paulo, S\~ao Carlos, 13560-970 S\~ao Paulo,
Brazil} 

\author{G. Vignale}
\affiliation{Department of Physics and Astronomy, University of
Missouri-Columbia, Columbia, Missouri, 65211}

\author{C. A. Ullrich}
\affiliation{Department of Physics and Astronomy, University of
Missouri-Columbia, Columbia, Missouri, 65211}

\date{\today }

\begin{abstract}
Energy gaps are crucial aspects of the electronic structure of finite and
extended systems. Whereas much is known about how to define and calculate
charge gaps in density-functional theory (DFT), and about the relation between
these gaps and derivative discontinuities of the exchange-correlation 
functional, much less is know about spin gaps. In this paper we give
density-functional definitions of spin-conserving gaps, spin-flip gaps and
the spin stiffness in terms of many-body energies and in terms of
single-particle (Kohn-Sham) energies. Our definitions are as analogous as
possible to those commonly made in the charge case, but important differences
between spin and charge gaps emerge already on the single-particle level
because unlike the fundamental charge gap spin gaps involve excited-state
energies. Kohn-Sham and many-body spin gaps are predicted to differ, and the
difference is related to derivative discontinuities that are similar to, but
distinct from, those usually considered in the case of charge gaps. Both
ensemble DFT and time-dependent DFT (TDDFT) can be used to calculate these
spin discontinuities from a suitable functional. We illustrate our findings
by evaluating our definitions for the Lithium atom, for which we calculate
spin gaps and spin discontinuities by making use of near-exact Kohn-Sham
eigenvalues and, independently, from the single-pole approximation to TDDFT.
The many-body corrections to the Kohn-Sham spin gaps are found to be negative,
{\em i.e.,} single-particle calculations tend to overestimate spin gaps while
they underestimate charge gaps.
\end{abstract}

\pacs{71.15.Mb, 31.15.ej, 31.15.ee, 72.25.Dc}


\newcommand{\be}{\begin{equation}}
\newcommand{\ee}{\end{equation}}
\newcommand{\bea}{\begin{eqnarray}}
\newcommand{\eea}{\end{eqnarray}}
\newcommand{\bi}{\bibitem}
\newcommand{\la}{\langle}
\newcommand{\ra}{\rangle}
\newcommand{\ua}{\uparrow}
\newcommand{\da}{\downarrow}
\renewcommand{\r}{({\bf r})}
\newcommand{\rp}{({\bf r'})}
\newcommand{\eps}{\epsilon}
\newcommand{\bfr}{{\bf r}}

\maketitle


\section{Introduction}

There is hardly any electronic property of a system that does not depend on
whether there is an energy gap for charge excitations, or for particle
addition and removal. Similarly, there is hardly any magnetic property of
a system that does not depend in some way on whether there is an energy gap
for flipping a spin, or for adding and removing spins from the system.

The reliable calculation of charge gaps\cite{footnote1} from first principles
is nontrivial and still faces practical problems (relevant aspects are
reviewed below), but at least conceptually it is clear how charge gaps are
to be defined and quantified within modern electronic-structure methods,
such as density-functional theory (DFT).\cite{dftbook,parryang,qtel}
On the other hand, much less is is known about how to calculate, or even
define, spin gaps.

In the present paper we show how to define and calculate the spin gap in
spin-DFT (SDFT), and predict that such calculations will encounter a spin-gap
problem similar to the band-gap problem familiar from applications of
DFT to semiconductors or to strongly correlated systems.

Section~\ref{chargegapsec} of this paper is devoted to charge gaps. In
Sec.~\ref{funexgaps} we recapitulate the conceptual difference between
fundamental gaps and excitation gaps. In Sec.~\ref{chargegapdef} we then
recall the quantitative definition of the fundamental gap and related
quantities, such as the single-particle gap, and particle addition and removal
energies. Section~\ref{chargediscsec} summarizes key aspects of the derivative
discontinuity, while Sec.~\ref{ensembleDFT} describes the connection between
gaps and discontinuities within the framework of ensemble DFT. Although the
final results of these sections are well known, our treatment is different
from the usual one in so far as we introduce many-body corrections to the gap
and derivative discontinuities in completely independent ways, related only
{\em a posteriori} via ensemble DFT. This way of proceeding is useful for
performing the generalization to the spin case.

For both the fundamental gap and the optical excitation
gap, the gapped degree of freedom is related to particles: either particles
are added to or removed from the system, or particles are excited to higher
energy levels within the system under study. In ordinary atoms, molecules and
solids, these particles are electrons, and the particle gaps of many-electron
systems are a key property in determining the functionality of today's
electronic devices.

The last decade has witnessed an enormous growth of interest in another type
of system, and in devices resulting from them, in which the key degree of
freedom is the spin. In the resulting field of spintronics, and the
development of spintronic devices, one is interested in controlling and
manipulating the spin degrees of freedom independently of, or in addition to,
the charge degrees of freedom. Here, the issue of the spin gap arises, and a
number of questions for electronic-structure and many-body theory appear:
What is the energy required to add a spin to the system? What is the energy
cost of flipping a spin? How do these concepts differ from the
fundamental and optical gaps involving particles? Can we calculate spin
gaps from spin-density-functional theory, and if yes, what type of
exchange-correlation ($xc$) functional is required? In Sec.~\ref{spingapsec},
we answer these questions.

In Sec.~\ref{spingapchargegap} we contrast spin gaps with charge gaps, and in
Sec.~\ref{spingapdef} we propose a set of many-body and single-particle
definitions for quantities related to the spin gap, such as spin-flip energies
and the spin stiffness. We take care to ensure that all quantities appearing
in our definitions can, in principle, be calculated from conventional SDFT or
time-dependent SDFT (TDSDFT), and try to make
the definitions in the spin case as analogous as possible to the charge case.
However, this analogy can only be carried up to a certain point, and important
differences between charge gaps and spin gaps emerge already at this level.
As a simple example, we consider, in Sec.~\ref{lisec}, the Lithium atom, for
which we confront calculated and experimental spin gaps.

In Sec.~\ref{spindiscsec} we then use ensembles DFT to relate the spin gap
to a derivative discontinuity that is similar to, but distinct from, the one
usually considered in the charge case. Finally, in Sec.~\ref{tddftsec}, we
investigate the connection to excitation gaps calculated from TDSDFT.
Equations are given that allow one to extract the various spin gaps and
related quantities from noncollinear TDSDFT calculations. For illustrative
purposes we evaluate these for the Lithium atom, and compare the gaps and
discontinuities obtained from time-dependent DFT to those obtained in
Sec.~\ref{lisec} from time-independent considerations.

Sec.~\ref{concl} contains our conclusions.

\section{Charge gap}
\label{chargegapsec}

\subsection{Fundamental gaps vs. excitation gaps}
\label{funexgaps}

To provide the background for this investigation, let us first briefly
recapitulate pertinent aspects of charge (or particle\cite{footnote1}) gaps.

While by definition all gaps involve energy differences between a lower-lying
state (in practice often the ground state) and a state of higher energy,
important differences depend on how the extra energy is added to the system
and what degrees of freedom absorb it. Therefore, different notions of gap
are appropriate for different purposes. For processes in which particles are
added to or removed from the system, which is subsequently allowed to relax
to the ground state appropriate to the new particle number, the key quantity
is the fundamental gap (sometimes also called the quasiparticle gap) which
is calculated from differences of
ground-state energies of systems with different particle number. As such, it
is relevant for instance in transport phenomena and electron-transfer reactions. If
energy is added by means of radiation, on the other hand, the particle number
does not change, and the relevant gap is an excitation energy of the
$N$-particle system. This excitation gap (sometimes also called the optical
gap), is relevant, e.g., in spectroscopy.

In first-principles electronic-structure calculations, excitation gaps are
today often calculated from time-dependent density-functional theory (TDDFT).
Fundamental gaps, on the other hand, involve ground-state energies of systems
with different particle numbers, and should thus, in principle, be accessible
by means of static (ground-state) DFT. However, it is well known that common
approximations to DFT encounter difficulties in this regard. In semiconductors,
for example, calculated fundamental gaps are often greatly underestimated
relative to experiment, and in strongly-correlated systems such as
transition-metal oxides, gapped materials are frequently incorrectly predicted
to be metallic, i.e., to have no gap at all. The resulting band-gap
problem of DFT has been intensely studied for many decades.

A major breakthrough in this field was the discovery of the derivative
discontinuity of the exact exchange-correlation ($xc$) functional of DFT,
which was shown to account for the difference between the gap obtained from
solving the single-particle Kohn-Sham (KS) equations of DFT, and the true
fundamental gap.\cite{pplb,pl,ss} The problems occurring in practice for
semiconductors and strongly-correlated systems are therefore attributed to
the fact that common local and semilocal approximations to the exact $xc$
functional do not have such a discontinuity. The development of DFT-based
methods allowing to nonempirically predict the presence and size of gaps in
many-electron systems continues to be a key issue of electronic-structure
theory and computational materials science.

\subsection{Definition of fundamental charge gaps}
\label{chargegapdef}

The {\em fundamental charge gap} $E_g$ is defined as the difference
\be
E_g(N) = I(N)-A(N),
\label{fungap}
\ee
where the electron affinity (energy gained by bringing in a particle from
infinity) and ionization energy (energy it costs to remove a particle to
infinity) are defined in terms of {\em ground-state energies}
of the $N$-particle system, as
\begin{eqnarray}
A(N)&=&E(N)-E(N+1)
\label{adef}
\\
I(N)&=&E(N-1)-E(N).
\label{idef}
\end{eqnarray}
The order of terms in these differences is the conventional choice.
The definition of the fundamental gap is in terms of processes involving
addition and removal of charge and spin. The change in the
respective quantum numbers is $\pm 1$ in $N$, and $\pm 1/2$ in $S$.
In chemistry,\cite{parryang} the average of $I$ and $A$ is identified with
the electronegativity of the $N$-particle system: $(I(N)+A(N))/2 = \chi(N)$.

The corresponding {\em Kohn-Sham gap} is defined analogously as
\be
E_{g,KS}(N) = I_{KS}(N)-A_{KS}(N),
\label{ksgap}
\ee
where $I_{KS}(N) = E_{KS}(N-1)-E_{KS}(N)$ and
$A_{KS}(N)=E_{KS}(N)-E_{KS}(N+1)$. Since the KS total energy is simply the sum
of the KS eigenvalues, $E_{KS}=\sum_{k=1}^N \eps_k$, this reduces to
$I_{KS}(N) = - \eps_N(N)$ and $A_{KS}(N) = - \eps_{N+1}(N)$, from which one
obtains the usual form
\be
E_{g,KS}(N) = \eps_{N+1}(N) - \eps_N(N),
\label{ksgap2}
\ee
where $\eps_N(N)$ and $\eps_{N+1}(N)$ are the highest occupied and the lowest
unoccupied state of the $N$-particle system, respectively.

The fundamental gap can also be written in terms of
KS eigenvalues by means of the ionization-potential theorem
(sometimes known as Koopmans' theorem of DFT), which states
\begin{eqnarray}
I(N)&=& -\epsilon_{N}(N)
\\
A(N) &=& I(N+1) = -\epsilon_{N+1}(N+1),
\end{eqnarray}
so that $I(N) \equiv I_{KS}(N)$, and
\be
E_g(N) = \epsilon_{N+1}(N+1) - \epsilon_{N}(N).
\label{fungap1}
\ee
Note that in contrast with the KS gap (\ref{ksgap2}) these eigenvalues
pertain to different systems.

The relation between both gaps is established by rewriting the fundamental
gap as
\be
E_g = E_{g,KS} + \Delta_{xc},
\label{deltaxcdef}
\ee
which defines $\Delta_{xc}$ as the $xc$ correction to the single-particle
gap. By making use of the previous relations we can cast $\Delta_{xc}$
as\cite{shamschlueter,seidl,confined,moldisc}
\be
\Delta_{xc} = \epsilon_{N+1}(N+1) - \eps_{N+1}(N) = A_{KS}(N) - A(N).
\label{discreps}
\ee
The important thing to notice in these expressions is that, due to protection
by Koopmans' theorem, {\em the ionization energy does not contribute to the
$xc$ correction $\Delta_{xc}$, so that the correction of the affinity and of
the fundamental gap are one and the same quantity.} Also, note that all of these
definitions can be made without any recourse to ensemble DFT and without
any mention of derivative discontinuities.

\subsection{Nonuniqueness and derivative discontinuities}
\label{chargediscsec}

The basic Euler equations of DFT is\cite{dftbook,parryang,qtel}
\begin{equation}
\frac{\delta E[n]}{\delta n\r} = \mu.
\label{euler}
\end{equation}
Since $E[n]=F[n]+\int d^3r \, v_{ext}\r n\r$ and
$E_{KS}[n]=T_s[n]+\int d^3r \, v_s\r n\r$, this implies
\begin{equation}
\frac{\delta F[n]}{\delta n\r} = \mu - v_{\rm ext}\r
\label{euler1}
\end{equation}
and
\begin{equation}
\frac{\delta T_s[n]}{\delta n\r} = \mu - v_s\r,
\label{euler2}
\end{equation}
where $T_s[n]$ is the noninteracting kinetic energy functional,
$F[n]= T[n]+U[n]$ is the internal energy functional, expressed in terms
of the interacting kinetic energy $T[n]$ and the interaction energy
$U[n]$, $\mu$ is the chemical potential, $v_{ext}\r$ is the external
potential and $v_s\r$ the effective KS potential.

Both the effective and the external potential are only defined up to a
constant, which does not change the form of the eigenfunctions. Consider now
a gapped open system, connected to a particle reservoir with fixed chemical
potential initially in the gap, and gradually change the constant. As long
as the change is sufficiently small, the chemical potential remains in the gap,
the density $n\r$ does not change, and the derivatives on the left-hand side
of Eqs.~(\ref{euler1}) and (\ref{euler2}) change continuously

However, once the change in the constant is large enough to affect the number
of occupied levels, the situation changes: As soon as a new level falls below
the chemical potential, or emerges above it, the number of particles in the
system changes discontinuously by an integer, and the chemical potential
adjusts itself to the new total particle number. For later convenience we call
the two values of $\mu$ on the left and the right of integer particle number
$\mu_-$ and $\mu_+$, respectively.

When the right-hand side of Eqs.~~(\ref{euler1}) and (\ref{euler2}) changes
discontinuously, the left-hand side must also change discontinuously.
This means that the functional derivatives of $F[n]$ and $T_s[n]$ change
discontinuously for variations $\delta n\r$ such that $N$ passes through
an integer, and are not defined precisely at the integer. We can also
argue conversely that if the functional derivatives existed at all $n\r$
they would determine the potentials uniquely. Since the potentials are
unique only up to a constant, the functional derivatives cannot exist
for the density variations $\delta n$ arising from changing the potential
by a constant. In a gapped system, these are the $\delta n$ integrating to
an integer.

Either way, we see that the indeterminacy of the potentials with respect to a
constant implies that the functionals $F[n]$ and $T_s[n]$ display derivative
discontinuities for certain directions in density space along which the
total particle number changes by an integer. This is the famous integer
discontinuity of DFT.\cite{pplb,pl,ss}

\subsection{Connection of discontinuities and gaps: ensemble DFT}
\label{ensembleDFT}

Up to this point we have defined $\Delta_{xc}$ as a many-body correction
to the single-particle gap, and deduced the existence of derivative
discontinuities from noting the nonuniqueness of the external potentials
with respect to a constant. These two conceptually
distinct phenomena are related by ensemble DFT for systems with fractional
particle number, describing open systems in contact with a particle
reservoir.\cite{pplb,pl,ss} For such systems ensemble DFT guarantees that
the ground-state energy as a function of particle number, $E(N)$, is a set
of straight lines connecting values at integer particle numbers.

For straight lines, the derivative at any $N$ can be obtained from the
values at the endpoints:
\begin{equation}
-A = E(N+1)-E(N)=\left.\frac{\partial E}{\partial N}\right|_{N+\delta N}
= \mu_+ = \left.\frac{\delta E}{\delta n\r}\right|_{N+\delta N}
\end{equation}
and
\begin{equation}
-I = E(N)-E(N-1)=\left.\frac{\partial E}{\partial N}\right|_{N-\delta N}
= \mu_-  = \left.\frac{\delta E}{\delta n\r}\right|_{N-\delta N}.
\end{equation}
The many-body fundamental gap is thus the derivative discontinuity of the
total energy across densities integrating to an integer:
\begin{equation}
E_g=I-A=\left.\frac{\delta E[n]}{\delta n\r}\right|_{N+\delta N}
- \left.\frac{\delta E[n]}{\delta n\r}\right|_{N-\delta N}.
\end{equation}

This energy functional is commonly written as $E=T_s + V + E_H + E_{xc}$,
where the external potential energy $V$ and the Hartree energy $E_H$ are
manifestly continuous functionals of the density. Hence, the energy gap
reduces to the sum of the discontinuity of the noninteracting kinetic energy
$T_s$ and that of the $xc$ energy $E_{xc}$.

The entire argument up to this point can be repeated for a noninteracting
system in external potential $v_s$. The energy of this system is
$E_{KS}=T_s+V_s$, of which only the first term can be discontinuous. Hence
the fundamental gap of the KS system is given by the discontinuity of $T_s$
\begin{equation}
E_{g,KS}=\left.\frac{\delta T_s[n]}{\delta n\r}\right|_{N+\delta N}
- \left.\frac{\delta T_s[n]}{\delta n\r}\right|_{N-\delta N}.
\end{equation}

Returning now to the many-body gap, written as the sum of the discontinuities
of $T_s$ and $E_{xc}$, we arrive at
\begin{equation}
E_g= E_{g,KS}+
\left.\frac{\delta E_{xc}[n]}{\delta n\r}\right|_{N+\delta N}
- \left.\frac{\delta E_{xc}[n]}{\delta n\r}\right|_{N-\delta N},
\end{equation}
or, by means of Eq.~(\ref{deltaxcdef}),
\begin{equation}
\Delta_{xc} = \left.\frac{\delta E_{xc}[n]}{\delta n\r}\right|_{N+\delta N}
- \left.\frac{\delta E_{xc}[n]}{\delta n\r}\right|_{N-\delta N}.
\label{dxcdef}
\end{equation}
This identifies the $xc$ correction to the single-particle gap as the
derivative discontinuity arising from the nonuniqueness of the
potentials with respect to an additive constant.\cite{pplb,pl,ss}

Importantly, {\em this connection is not required to define the $xc$
corrections} and neither is its existence enough to conclude that these
corrections are nonzero. Many-body corrections to the single-particle gap
can be defined independently of any particular property of the
density functional (or even without using any density-functional theory),
and whether for a given system these corrections are nonzero or not
depends on the electronic structure of that particular system, and does not
follow from the formal possibility of a derivative discontinuity, because this
discontinuity itself might be zero. Thus, the question of the existence and
size of $xc$ corrections to the charge gap must be asked for each system anew.
As we will see in the next section, the same is true for the spin gap.

\section{Spin gap}
\label{spingapsec}

We have provided the above rather detailed summary of the definition of the
fundamental charge gap and its connection to nonuniqueness and to derivative
discontinuities to prepare the ground for the following discussion of the spin
gap. In order to arrive at a consistent DFT definition of spin gaps, we follow
the steps outlined in the charge case: (i) define appropriate gaps and their
$xc$ corrections, (ii) use the nonuniqueness of the SDFT potentials to show
the existence of spin derivative discontinuities, and (iii) identify a
suitable spin ensemble to connect the two.

\subsection{Spin gap vs. charge gap}
\label{spingapchargegap}

To introduce a spin gap or a spin-flip energy (see below for precise
definitions) we consider processes in which only the total spin of the system
is changed, while the particle number remains the same. There cannot be
any definition in terms of particle addition and removal energies, since
in these processes the charge changes, too, which is not what one wants
the spin gap to describe.
In other words, the change of quantum numbers related to a
spin flip is $\pm 1$ for the spin  and $0$ for charge. Note
that this is an excitation energy, where the excitation takes place under the
constraints of constant particle number and change of total spin by one unit.
This is the key difference to the previous section, from which all other
differences follow.

\subsection{Definition of spin gaps: spin-flip energies and spin stiffness}
\label{spingapdef}

First, we define the {\em spin up-flip energy} and the {\em spin down-flip
energy} in terms of many-body energies as
\begin{eqnarray}
E^{sf+}(N) = E(N,S+1)-E(N,S)
\label{mbsfup}
\\
E^{sf-}(N) = E(N,S-1)-E(N,S).
\label{mbsfdn}
\end{eqnarray}
Here $E(N,S)$ is the lowest energy in the $N$-particle spin-$S$ subspace,
where $S$ is the eigenvalue of the $z$-component of the total spin, and we
assumed that spin-up and spin-down are good quantum numbers. This implies,
in particular, that spin-orbit coupling is excluded from our
analysis. (Of course these definitions only apply if the respective flips are
actually possible; in other words, if $S$ does not yet have the maximal or
minimal value for a given $N$.)

The differences $E^{sf+}(N)$ and $E^{sf-}(N)$
are similar to the concepts of affinity and ionization
energy, Eqs.~(\ref{adef}) and (\ref{idef}). However, affinities and
ionization energies are always defined with the smaller value (of $N$) as
the first term in the differences, whereas spin-flip energies are
conventionally defined as final state minus initial state, i.e. both
spin-flip energies measure an energy cost. Therefore, the down-flip is the
spin counterpart to the ionization energy, while the up-flip is the spin
counterpart to {\em minus} the electron affinity.

A more important difference is that the spin-flip energies involve excited-state
energies $E(N,S+1)$ and $E(N,S-1)$ of the $N$-particle system, instead
of ground-state energies, and in this sense are more similar to the optical
gap in the charge case than to the quantities used in evaluating the
fundamental gap. Alternatively, these energies can also be considered
ground-state energies for sectors of Hilbert space restricted to a given
total $S$, but we will not make use of this alternative interpretation in the following.

\begin{figure}
\centering
\includegraphics[width=\linewidth]{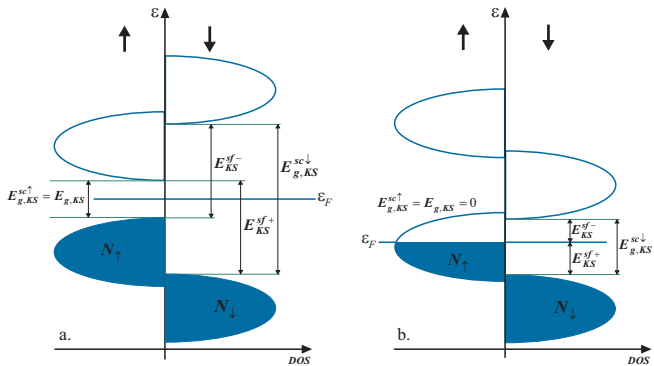}
\caption {\label{fig1}
Left: Spin-resolved single-particle (KS) density-of-states of a spin-polarized
insulator. Two spin-flip energies and two spin-conserving gaps can be defined.
Right: The half-metallic ferromagnet is a special case in which the
gap in one spin channel (say spin up) is zero. In this case, there is only
one spin-conserving gap, equal to the sum of both spin-flip energies,
$E_{KS}^{sf-} + E_{KS}^{sf+} = E_{s,KS}^{sc}$, and the KS charge gap is
zero, due to the presence of the gapless spin down channel.
Figure courtesy of Daniel Vieira.}
\end{figure}

KS spin-flip energies are related analogously to single-particle eigenvalues,
according to
\bea
E_{KS}^{sf+} &=&\epsilon_{l(\uparrow)}-\epsilon_{H(\downarrow)}
\label{ksspinflipdefs1}
\\
E_{KS}^{sf-}&=&\epsilon_{l(\downarrow)}-\epsilon_{H(\uparrow)},
\label{ksspinflipdefs2}
\eea
where all energies are calculated at the same $N,S$.
Here $l(\sigma)$ means the lowest unoccupied spin $\sigma$ state, and
$H(\sigma)$ means the highest occupied spin $\sigma$ state. Similarly,
$L(\sigma)$ and $h(\sigma)$ denote the lowest occupied spin $\sigma$
state and $h(\sigma)$ the highest unoccupied spin $\sigma$ state,
respectively. This notation is nonstandard, but helpful, and further
illustrated in Fig.~\ref{fig1}.

In the same way, we can also define the spin conserving ({\em sc}) single-particle
gaps in each spin channel, as
\bea
E_{g,KS}^{sc,\uparrow} &=&\epsilon_{l(\uparrow)}-\epsilon_{H(\uparrow)}
\\
E_{g,KS}^{sc,\downarrow}&=&\epsilon_{l(\downarrow)}-\epsilon_{H(\downarrow)}.
\label{ksscgapdefs}
\eea
The spin-conserving gaps and the spin-flip energies are necessarily related by
$E_{g,KS}^{sc\uparrow}+E_{g,KS}^{sc\downarrow}
=E_{KS}^{sf-}+E_{KS}^{sf+}$.
A look at Fig.~\ref{fig1} clarifies these definitions. If the system
is non spin polarized, both spin-flip energies and both spin-conserving
gaps become equal to the ordinary KS charge gap, which in our present
notation reads $\eps_l-\eps_H$.

In the same way as for charge gaps, we can now also consider the sum and the
difference of the spin-flip energies. The sum
\be
E_s = E^{sf-} + E^{sf+}
\label{spinstiffness}
\ee
of the energies it costs to flip a spin up and a spin down is formally
analogous to the fundamental gap (\ref{fungap}), but with the important
difference that $E_s$ involves excited-state energies.
The unusual sign (sum instead of difference) arises simply because both
spin-flip energies measure costs, whereas the affinity featured in
Eq.~(\ref{fungap}) measures an energy gain.

The formal analogy to Eq.~(\ref{fungap}) suggests that the quantity
defined in Eq.~(\ref{spinstiffness})
be called the fundamental spin gap. In practice, however, the name spin gap
is more appropriately applied to the individual spin-flip energies. The
physical interpretation of their sum, Eq.~(\ref{spinstiffness}), is revealed
by expressing it in terms of the many-body energies by means of Eqs.
(\ref{mbsfup}) and (\ref{mbsfdn}):
\be
E_s =E(N,S+1) + E(N,S-1) - 2 E(N,S).
\label{spinstiffness2}
\ee
This is of the form of a discretized second derivative
$\partial^2 E(N,S)/\partial S^2$, which identifies $E_s$ as the
discretized {\em spin stiffness} [we anticipated this interpretation
when attaching a subscript $s$ for stiffness to the sum in
Eq.~(\ref{spinstiffness})]. We note that half of $I-A$ is known in
quantum chemistry as {\em chemical hardness}, which conveys a very
similar idea as stiffness. Generically, we refer to all three quantities
$E^{sf-}$, $E^{sf+}$ and $E_s$ as spin gaps.

The {\em spin electronegativity} can be defined as half of the difference of the
spin-flip energies, $\chi^s = (E^{sf-} - E^{sf+})/2$.
This quantity has the following interpretation:
if $\chi^s>0$, it costs less energy to flip a spin up than to flip a spin
down, whereas if $\chi^s<0$ the down flip is energetically cheaper.


The KS spin stiffness is defined as the sum of KS spin-flip energies,
\be
E_{s,KS} = E_{KS}^{sf-} + E_{KS}^{sf+},
\label{ksspingapdef}
\ee
or, with Eqs.~(\ref{ksspinflipdefs1}) and (\ref{ksspinflipdefs2}),
\be
E_{s,KS} = [\epsilon_{l(\downarrow)}-\epsilon_{H(\uparrow)}]
+ [\epsilon_{l(\uparrow)} - \epsilon_{H(\downarrow)}].
\label{ksspinstiffness}
\ee
This is analogous to (\ref{ksgap2}), except that in spin flips nothing is
removed to infinity or brought in from infinity. Thus, differently from the
KS ionization energy and electron affinity, the spin-flip energies require two single-particle
energies for their definition instead of one, and in contrast with
the KS charge gap the KS spin stiffness requires four single-particle energies instead of two.

This missing analogy is physically meaningful:
Conventional gaps are defined in terms of particle addition and removal
processes and are ground-state properties. To define pure spin gaps
(i.e., spin-flip energies and spin stiffness)
in which the charge does not change, we cannot make use of particle
addition and removal processes but have to use spin flip processes
instead. However, {\em spin flips are excitation energies, and we must
specify both initial and final states to define them properly.}

We also note that the many-body spin stiffness has no simple expression in
terms of eigenvalues which would be analogous to Eq.~(\ref{fungap1}). Such
an expression would require the spin counterpart to Koopmans' theorem
$I(N) \equiv I_{KS}(N)$, which is not available for spin-flip energies.
Hence, in general both spin-flip energies $E^{sf-}$ and $E^{sf+}$ may be
individually different from their KS counterparts $E_{KS}^{sf-}$ and
$E_{KS}^{sf+}$:
\bea
E^{sf-} &=& E_{KS}^{sf-} + \Delta_{xc}^{sf-} \label{deltaplus}
\\
E^{sf+} &=& E_{KS}^{sf+} + \Delta_{xc}^{sf+}. \label{deltaminus}
\eea
We can, moreover, establish a relation between the many-body spin stiffness
and the KS spin stiffness by rewriting the former as
\be \label{delta}
E_s = E_{s,KS} + \Delta_{xc}^s
=E_{KS}^{sf-} + E_{KS}^{sf+} + \Delta_{xc}^s,
\label{spingap2}
\ee
which defines $\Delta_{xc}^s$ as the $xc$ correction to the
KS spin stiffness.
The important thing to notice in Eq.~(\ref{spingap2}) is that {\em there is
no reason to attribute $\Delta_{xc}^s$ only to the up-flip energy.} This is
a key difference to the charge case, where $I(N) \equiv I_{KS}(N)$
and the $xc$ correction could thus be attributed only to the electron
affinity. Rather, the spin-flip corrections are connected by
\be
\Delta_{xc}^s = \Delta_{xc}^{sf-} + \Delta_{xc}^{sf+}. \label{delta1}
\ee

\subsection{Example: the Li atom}
\label{lisec}

\begin{table}
\caption{\label{table1} Kohn-Sham energy eigenvalues (in eV) for the Lithium atom. The
$1s\ua$, $2s\ua$ and $1s\da$ levels are occupied. KS: energy eigenvalues obtained by
inversion from quasi-exact densities. XX denotes exact exchange,\cite{krieger}
and KLI is the Krieger-Li-Iafrate approximation.\cite{kli}}
\begin{ruledtabular}
\begin{tabular}{lccccc}
  & KS\footnote{Ref. \onlinecite{krieger}} & KS\footnote{Ref. \onlinecite{gritsenko}}&
 XX\footnotemark[1] & KLI-XX & LSDA  \\ \hline
$-\epsilon_{1s\ua}$  & 55.97 & 58.64 & 55.94  & 56.64 & 51.02\\
$-\epsilon_{2s\ua}$  & 5.39  & 5.39  & 5.34   &  5.34 & 3.16\\
$-\epsilon_{2p\ua}$  & 3.54  &  -    & 3.48   &  3.50 & 1.34 \\
$-\epsilon_{1s\da}$  & 64.41 & 64.41 & 67.18  & 67.14 & 50.81\\
$-\epsilon_{2s\da}$  &  8.16 & 5.87  & 8.25   & 8.23  & 2.09
\end{tabular}
\end{ruledtabular}
\end{table}

\begin{table}
\caption{\label{table2} Single-particle spin-flip energies (\ref{Liplus})
and (\ref{Liminus}) and spin stiffness (\ref{ksspinstiffness}), their
experimental (Exp) counterparts, (\ref{mbsfup}), (\ref{mbsfdn}) and
(\ref{spinstiffness}), and the resulting {\em xc} corrections defined in
(\ref{deltaplus}), (\ref{deltaminus}) and (\ref{delta}), for the Lithium
atom. In the columns labelled KS we employ KS eigenvalues obtained
from near-exact densities, while in the columns labelled XX, KLI-XX and LSDA
we use approximate eigenvalues obtained from standard SDFT calculations.
The experimental values were obtained using spectroscopic data for the
lowest quartet state $^4P^0$ from Ref.~\onlinecite{Mannervik1983} as well as
accurate wave-function based theory from Ref.~\onlinecite{Hsu1991}.
All values are in eV. }
\begin{ruledtabular}
\begin{tabular}{lcccccc}
  & KS\footnote{Ref.~\onlinecite{krieger}} & KS\footnote{Ref.~\onlinecite{gritsenko},
  taking $-\epsilon_{2p\ua}=3.54$ eV from Ref.~\onlinecite{krieger}}&
 XX\footnotemark[1] &KLI-XX& LSDA & Exp \\ \hline
$E^{sf+}$   & 60.87    & 60.87 &  63.70 & 63.64 & 49.47 & 57.41 \\
$E^{sf-}$   & $-2.77$& $-0.48$  &  $-2.91$ & $-2.89$ & 1.07 & 0\\
$E_s$       & 58.10  &   60.39   &  60.79 & 60.75 & 50.54 & 57.41 \\
& & & & \\
$\Delta_{xc}^{sf+}$ & $-3.46$ & $-3.46$ & $-6.29$ & $-6.23$ & 7.94 &  \\
$\Delta_{xc}^{sf-}$ &  2.77   & 0.48   & 2.91 & 2.89 & $-1.07$ & \\
$\Delta_{xc}^s$     & $-0.69$ &  $-2.98$& $-3.38$ & $-3.34$ & 6.87 &
\end{tabular}
\end{ruledtabular}
\end{table}


To give an explicit example of the quantities introduced in the previous
section, we now consider the Li atom. For this system, KS eigenvalues
$\eps_{H\ua}=\eps_{2s\ua}$,
$\eps_{l\ua}=\eps_{2p\ua}$,
$\eps_{l\da}=\eps_{2s\da}$ and
$\eps_{H\da}=\eps_{1s\da}$ have been obtained by
numerical inversion of the KS equation starting from near-exact densities (see Table \ref{table1}).\cite{krieger,gritsenko}

The KS spin-flip energies are obtained as
\begin{eqnarray}
E_{KS}^{sf+} &=& \epsilon_{2p\ua} - \epsilon_{1s\da} \label{Liplus}\\
E_{KS}^{sf-} &=& \epsilon_{2s\da} - \epsilon_{2s\ua} \label{Liminus} .
\end{eqnarray}
They are given in Table \ref{table2}, together with the spin stiffness $E_s$, see Eq. (\ref{spinstiffness}).
Table \ref{table2} also presents the corresponding experimental many-body energy differences for the Li atom, which
were obtained using spectroscopic data for the lowest quartet state $^4P^0$ and
accurate wave-function based theory.\cite{Mannervik1983,Hsu1991}
Relativistic effects and other small corrections
included in the experimental data are ignored since they are too small on the
scale of energies we are interested in.

Table \ref{table2} also gives the $xc$ corrections to the single-particle spin
flip energies, see Eqs. (\ref{deltaplus}) and (\ref{deltaminus}), and the $xc$
correction to the spin stiffness, $\Delta_{xc}^s$, see Eq. (\ref{delta}). As
a consistency test we verified that relation (\ref{delta1}), which connects
the $xc$ corrections of the spin-flip energies to the $xc$ corrections of the
spin stiffness, is satisfied.

We also carried out calculations using the exact-exchange (XX) eigenvalues of
Ref.~\onlinecite{krieger} in order to separately assess the size of
exchange and correlation effects. The resulting value of
$\Delta_x^{s} = -3.38$ eV indicates a larger (more negative) correction than in the calculation
including correlation. An approximate KLI-XX calculation\cite{kli} yields very similar results,
while the LSDA data are completely different and do not even
reproduce the correct sign.

Three of the required ``exact'' KS single-particle eigenvalues are also reported in
Ref.~\onlinecite{gritsenko} (we use the result of Ref. \onlinecite{krieger} for the missing value of $\eps_{2p\ua}$).
The value of $\eps_{2s\da}$ is quite different than the
value reported in Ref. \onlinecite{krieger} ($-5.87$ eV versus $-8.16$ eV), and consequently
we obtain a rather different value of $\Delta_{xc}^{s}$ ($-2.98$ eV versus $-0.69$ eV).
Nevertheless, both sets of data sustain our main conclusions in this section:

(i) Simple LSDA calculations give rise to serious qualitative errors. As can be seen from Table \ref{table2},
one obtains spin-flip energies that are drastically to small ($E^{sf+}$) or have the wrong sign ($E^{sf-}$).
The resulting $xc$ corrections also suffer from having the wrong sign. These shortcomings of the LSDA are
hardly surprising in view of its well-established failure to describe the charge gap.

(ii) Even the precise KS eigenvalues do not
predict the exact spin flip energies and spin stiffness, i.e. the $xc$
corrections introduced in Sec.~\ref{spingapdef} on purely formal grounds
are indeed nonzero. The absolute size of these corrections implies that
a simple KS eigenvalue calculation of spin gaps can be seriously in error.

(iii) Exchange-only calculations overestimate (in modulus) the size of the
gap corrections. This implies that there is substantial cancellation between the
exchange and the correlation contribution to the full correction. This is
the same trend known for charge gaps.

(iv) The $xc$ corrections to both the up-flip energy and the spin stiffness
turn out to be negative; in other words the KS calculation overestimates
these quantities. This is the opposite of what occurs in the case of the
fundamental charge gap, which is underestimated by the KS calculation.
We note that hints of an underestimation of the experimental spin-flip
energies by KS eigenvalue differences have also been observed for
half-metallic ferromagnets. In the case of $\rm CrO_2$, for example,
Ref.~\onlinecite{coey} reports experimental spin-flip energies in the range
0.06 to 0.25 eV and compiles SDFT predictions that range from 0.2 to 0.7 eV
(and in one case even 1.7 eV).

\subsection{Nonuniqueness and derivative discontinuities in SDFT}
\label{spindiscsec}

Above we pointed out that the effective and external potentials of DFT
are determined by the ground-state density up to an additive constant.
However, this statement only holds when one formulates DFT exclusively in
terms of the charge density, as we have done in discussing charge gaps.
It does not hold when one works with spin densities, as in SDFT, or current
densities, as in current-DFT (CDFT).

In these cases the densities still determine the wave
function, but they do not uniquely determine the corresponding potentials.
A first example of this {\it nonuniqueness problem} of generalized DFTs
was already encountered in early work on SDFT, for the single-particle KS
Hamiltonian.\cite{vbh} Later, this observation was extended to the SDFT
many-body Hamiltonian,\cite{nonunep,nonun1} and further examples were
obtained in CDFT\cite{nonun2} and DFT on lattices.\cite{carstennonun}

Nonuniqueness is a generic feature of generalized (multidensity) DFTs,
consequences of which are still under investigation.\cite{stronghk,argaman,nikitas,kohnnonun,degnonun,gal}
In particular, Refs.~\onlinecite{nonunep} and \onlinecite{nonun1}
already point out that the nonuniqueness of the potentials of SDFT implies
that the SDFT functionals can have additional derivative discontinuities,
because, if the functional derivatives of $F$ and $T_s$ in multi-density DFTs
such as SDFT and CDFT existed for all densities, they would determine the
corresponding potentials uniquely. Very recently, G\'al and collaborators
\cite{gal} pointed out that one-sided derivatives may still exist, and
explored consequences of this for the DFT description of chemical reactivity
indices.

Just as in the charge case, derivative discontinuities result from the
nonuniqueness of the spin-dependent potentials, while corrections to
single-particle gaps result from the auxiliary nature of Kohn-Sham eigenvalues.
In the charge case, both distinct phenomena could be connected by means of
ensemble DFT for systems of fractional particle number. The question then
arises if a similar connection can also be established in the spin case.
This requires an investigation of spin-ensembles.

\subsection{Spin ensembles}
\label{spinensembles}

Consequences of the nonuniqueness of the potentials of SDFT for the
calculation of spin gaps were already hinted at in
Refs.~\onlinecite{nonunep} and \onlinecite{nonun1}, where it was pointed
out that there may be a spin-gap problem in SDFT similarly to the well
known band-gap problem of DFT.

To make these hints more precise, we first recall, from the above,
that the quantity usually called the spin gap is actually what we here
called the spin-flip gap, and is analogous to the ionization energy or the
electron affinity in the charge case, not to the fundamental particle gap.
The spin-dependent quantity that is most analogous to the fundamental particle
gap is the discretized spin stiffness of Eqs.~(\ref{spinstiffness}) and
(\ref{spinstiffness2}). However, regardless of whether one focuses on the
spin-flip energies or on the spin stiffness, the spin situation is not
completely analogous to the charge situation because
both the spin-flip gaps and the spin stiffness are defined in terms
of excited states of an $N$-particle Hamiltonian, while charge gaps are
defined in terms of ground-state energies of Hamiltonians with different
particle numbers.

To identify a suitable ensemble, we write the energy associated with a
generic ensemble of two systems, A and B, as
\be
E_w = (1-w) E_A + w E_B,
\ee
where $0 \leq w \leq 1$ is the ensemble weight.
If A and B have different particle numbers, $N_A$ and $N_B = N_A \pm 1$, this
becomes the usual fractional-particle number ensemble, which is
unsuitable for our present investigation where the involved systems differ
in the spin but not the charge quantum numbers.

A spin-dependent ensemble was recently
constructed by Yang and collaborators \cite{yang1,yang2} in order to
understand the static correlation error of common density functionals.
In this spin ensemble, A and B have different (possibly fractional) spin,
but are degenerate in energy. The constancy condition, whose importance and
utility is stressed in Refs.~\onlinecite{yang1,yang2}, arises directly from
the restriction of the ensemble to degenerate states. While useful for the
purposes of analyzing the static correlation error, this spin ensemble is
too restrictive for our purposes, as it excludes the excited states involved
in the definition of spin-flip gaps and of the spin stiffness.

Ensembles involving excited states have been employed in DFT in connection
with the calculation of excitation energies.\cite{teophilou,ogk,nagy}
Here A and B differ in energy but stem from the same Hamiltonian, with fixed
particle number. Excited-state ensemble theory leads to a simple expression
relating the first excitation energy to a KS eigenvalue difference\cite{ogk}
\be \label{ogk_eq}
E_B - E_A = \eps^w_{M+1} - \eps^w_M + \left. \frac{\partial E_{xc}^w[n]}{\partial w}\right|_{n=n_w},
\ee
where $E_B$ and $E_A$ are the energies of first excited and the ground state
of the many-body system, respectively, $\eps^w_{M+1}$ and $\eps^w_M$ are
the highest occupied and lowest unoccupied KS eigenvalues, and
$E^w_{xc}$ is the ensemble $xc$ functional. Equation (\ref{ogk_eq}) holds for ensemble weights in the
range $0 \le w \le 1/2$.
Levy showed \cite{levy} that the last term in this equation is related to
a derivative discontinuity according to
\be \label{levy_eq}
\left. \frac{\partial E_{xc}^w[n]}{\partial w}\right|_{n=n_w} =
\left.\frac{\delta E_{xc}^{w=0}[n]}{\delta n\r}\right|_{n = n^{w=0}}
- \left. \frac{\delta E_{xc}^w[n]}{\delta n\r}\right|_{n = n^ w}
\ee
for $w\to 0$. Here $n^w=(1-w)n_A + w n_B$ is the ensemble density, and the
discontinuity arises because even in the $w\to 0$ limit the ensemble density
does contain an admixture of the state B with energy $E_B > E_A$ and thus
decays differently from $n_0$ as $r\to \infty$.\cite{levy} Levy developed
his argument explicitly only for the spin-unpolarized case, but already
pointed out in the original paper that the results carry over to
spin-polarized situations.

In our case, we take A to be the ground state and B to be the lowest-lying
state differing from it by a spin flip. To be specific, let us assume that
the spin is flipped up. In this case we obtain from Eqs. (\ref{ogk_eq}) and (\ref{levy_eq}) in the limit $w\to 0$,
and using our present notation,
\bea
E^{sf+}(N)&=& E(N,S+1)-E(N,S) \\
&=&
\eps^w_{l(\ua)} - \eps^w_{H(\da)}
+ \left. \frac{\partial E_{xc}^w[n_\ua,n_\da]}{\partial w}\right|_{{n_\ua = n_\ua^w}\atop{n_\da=n_\da^w}} \label{ensembledisca}\\
&=&
\eps^w_{l(\ua)} - \eps^w_{H(\da)} +
\left.\frac{\delta E_{xc}^{w=0}[n_\ua,n_\da]}{\delta n_\da\r}
\right|_{{n_\ua = n_\ua^{w=0}}\atop{n_\da = n_\da^{w=0}}} \nonumber\\
&&
- \left. \frac{\delta E_{xc}^w[n_\ua,n_\da]}{\delta n_\da\r}
\right|_{{n_\ua = n_\ua^w}\atop{n_\da = n_\da^w}}
\\
&=& E^{sf+}_{w,KS}(N) + \Delta^{sf+}_{w,xc}
\label{ensembledisc}
\eea
for $w\to 0$. Equation (\ref{ensembledisc}), which is the ensemble version of 
our Eq.~(\ref{deltaminus}), illustrates that KS spin-flip excitations, too,
acquire a many-body correction arising from a derivative discontinuity.

In the particular case in which the spin flip costs no energy in the
many-body and in the KS system, the preceding equation reduces to
$\Delta^{sf+}_{w,xc}=0$, which is the constancy condition derived in
Refs.~\onlinecite{yang1,yang2} for spin ensembles of degenerate states.

We note that the KS eigenvalues and the discontinuity in
Eqs.~(\ref{ensembledisca}) to (\ref{ensembledisc}) must be evaluated by 
taking the $w \to 0$ limit of the $w$-dependent quantities, while the
quantities in Eq.~(\ref{deltaminus}) have no ensemble dependence. This
complicates the evaluation of spin-flip energies and their discontinuities,
as defined in Sec.~\ref{spingapdef}, from ensemble DFT.
Therefore, we turn to still another density-functional approach to excited
states in order to evaluate these quantities: TDDFT.

\subsection{Connection to TDDFT}
\label{tddftsec}

TDDFT has established itself
as the method of choice for calculating excitation energies in atomic and molecular systems,
and is making rapid progress in nanoscale systems and solids as well.\cite{tddftbook,Elliott2008}
In this section we will make a connection between the preceding discussion and TDDFT,
which will allow us to derive simple approximations for the $xc$ corrections to the single-particle
spin-flip excitation energies and the spin stiffness.


To calculate the spin-conserving {\em and} the spin-flip excitation energies, it is necessary
to use a {\em noncollinear} spin-density response theory, even if the system under study
has a ground state with collinear spins ({\em i.e.,} spin-up and -down with respect to the $z$ axis
are good quantum numbers). In this way the spin-up and spin-down density responses can become coupled,
and the description of spin-flip excitations (for instance, due to a transverse magnetic perturbation)
becomes possible. In TDDFT, the spin-conserving and the spin-flip excitation energies can be obtained from
the following eigenvalue equations, which are a generalization
of the widely used Casida equations\cite{Casida1996}
for systems with noncollinear spin:\cite{Ziegler}
\begin{widetext}
\begin{eqnarray} \label{C1}
\sum_{\sigma \sigma'}\sum_{i'a'}\left\{ \Big[
\delta_{i'i}\delta_{a'a} \delta_{\sigma\alpha}\delta_{\sigma'\alpha'} \omega_{a'\sigma'i'\sigma }
+ K_{i\alpha a\alpha',i'\sigma a'\sigma'}^{\alpha\alpha',\sigma\sigma'}
\Big]X_{i'\sigma a'\sigma'}
+
K_{i\alpha a\alpha', i'\sigma a'\sigma'}^{\alpha\alpha',\sigma'\sigma} Y_{i'\sigma,a'\sigma' }
\right\}
&=&
-\omega X_{i\alpha,a\alpha'}
\\
\sum_{\sigma \sigma'}\sum_{i'a'} \left\{
K_{ i\alpha a\alpha',i'\sigma a'\sigma'}^{\alpha'\alpha,\sigma\sigma'} X_{i'\sigma a'\sigma'}
+
\Big[
\delta_{a'a}\delta_{i'i} \delta_{\sigma'\alpha'}\delta_{\sigma\alpha} \omega_{a'\sigma' i'\sigma}
+ K_{ i\alpha a\alpha', i'\sigma a'\sigma'}^{\alpha'\alpha,\sigma'\sigma}
\Big]Y_{i'\sigma,a'\sigma' }\right\}
&=&
\omega Y_{i\alpha,a\alpha'} \:,\label{C2}
\end{eqnarray}
\end{widetext}
where we use the standard convention that $i,i$
and $a,a'$ are indices of occupied and unoccupied KS orbitals, respectively, and $\alpha' \alpha', \sigma\sigma'$ are spin indices,
and $\omega_{a'\sigma'i'\sigma } = \epsilon_{a'\sigma'} -\epsilon_{i'\sigma}$.
Choosing the KS orbitals to be real, without loss of generality, we have
\begin{eqnarray}
\hspace*{-4mm}
K_{i\alpha a\alpha',i'\sigma a'\sigma'}^{\alpha\alpha',\sigma\sigma'}(\omega) &=&
\int\! d^3r \! \int\! d^3r' \: \psi_{i\alpha}({\bf r}) \psi_{a\alpha'}({\bf r})
\nonumber\\
&\times&
 f^{Hxc}_{\alpha\alpha',\sigma\sigma'}(\bfr,\bfr',\omega) \psi_{i'\sigma}({\bf r}') \psi_{a'\sigma'}({\bf r}') \:.
\end{eqnarray}
Here, the subscript indices of the matrix elements $K$ refer to the KS orbitals in the integrand,
and the superscript spin indices refer to the Hartree-$xc$ kernel
\begin{equation}
f^{Hxc}_{\alpha,\alpha',\sigma\sigma'}(\bfr,\bfr',\omega)
=
\frac{\delta_{\alpha\alpha'}\delta_{\sigma \sigma'}}{|\bfr - \bfr'|} +
f^{xc}_{\alpha \alpha',\sigma\sigma'}(\bfr,\bfr',\omega) \:,
\end{equation}
where the frequency-dependent $xc$ kernel is defined as the Fourier transform of the time-dependent $xc$ kernel
\begin{equation}
f^{xc}_{\alpha \alpha',\sigma\sigma'}(\bfr,t,\bfr',t') = \left.
\frac{\delta v^{xc}_{\alpha\alpha'}(\bfr,t)}{\delta n_{\sigma\sigma'}(\bfr',t')}
\right|_{\underline{n}(\bfr,t) = \underline{n}^0(\bfr)}.
\end{equation}
Here, $\underline{n}(\bfr,t)$ and $\underline{n}^0(\bfr)$ are the time-dependent and the ground-state $2\times 2$
spin-density matrix, which follow from the DFT formalism
for noncollinear spins. \cite{Ziegler,kubler,sandratskii,heinonen,UllrichFlatte}

Eqs. (\ref{C1}) and (\ref{C2}) give, in principle, the exact spin-conserving and spin-flip excitation energies of the
system, provided the exact KS orbitals and energy eigenvalues are known, as well as the exact functional form of
$f^{xc}_{\alpha \alpha',\sigma \sigma'}$. We will now consider a simplified solution known as the
single-pole approximation. \cite{Petersilka1996,Appel2003} It is obtained from the full system of equations (\ref{C1}), (\ref{C2})
by making the Tamm-Dancoff approximation (i.e., neglecting the off-diagonals) and focusing only on the $H (\sigma)\to l(\sigma')$
excitations. In other words, we need to solve the $4\times 4$ problem
\begin{equation}
\sum_{\sigma \sigma'}
\Big[ \delta_{\sigma'\alpha'}\delta_{\sigma\alpha} \omega_{l\sigma' H\sigma}
+ K_{ H\alpha l\alpha', H\sigma l\sigma'}^{\alpha'\alpha,\sigma'\sigma}
\Big]Y_{H\sigma,l\sigma' }
=
\omega Y_{H\alpha,l\alpha'} \:.
\end{equation}

For ground states with collinear spins, the only nonvanishing elements of the Hartree-$xc$ kernel are
\begin{equation}
f^{Hxc}_{\ua \ua,\ua \ua} , \;
f^{Hxc}_{\da \da,\da \da} , \;
f^{Hxc}_{\ua \ua,\da \da} , \;
f^{Hxc}_{\da \da,\ua \ua} , \;
f^{xc}_{\ua \da,\ua \da} , \;
f^{xc}_{\da \ua,\da \ua}
\end{equation}
(notice that there is no Hartree term in the spin-flip channel),
and the spin-conserving and spin-flip excitation channels decouple into two separate $2\times 2$ problems.
For the spin-conserving case, we have
\begin{equation}
\det \left|
\begin{array}{cc}
\omega_{\ua \ua} - \omega^{sc} + M_{\ua \ua,\ua \ua} & M_{\ua \ua,\da ,\da}\\
M_{\da \da,\ua \ua} & \omega_{\da \da} - \omega^{sc} + M_{\da \da,\da \da}
\end{array}
\right|=0 \:,
\end{equation}
where we abbreviate $M_{\alpha \alpha',\sigma\sigma'} = K_{ H\alpha l\alpha', H\sigma l\sigma'}^{\alpha'\alpha,\sigma'\sigma}(\omega)$
and  $\omega_{\sigma'\sigma} = \omega_{l\sigma',H\sigma} = \epsilon_{l\sigma'} - \epsilon_{H\sigma}$.
From this, we get the two spin-conserving excitation energies as
\begin{eqnarray} \label{omegasc}
E^{sc\ua, \da}&=& \frac{\omega_{\ua\ua} + \omega_{\da\da} + M_{\ua \ua,\ua \ua} + M_{\da \da,\da \da}}{2}
\pm \Big[M_{\ua \ua,\da \da}M_{\da \da,\ua \ua}
\nonumber\\
&+&
\frac{1}{4}(\omega_{\ua \ua} - \omega_{\da \da} + M_{\ua \ua,\ua \ua} - M_{\da \da,\da \da})^2
\Big]^{1/2},
\end{eqnarray}
with the spin-conserving Kohn-Sham single-particle gaps $E^{sc\ua(\da)}_{g,KS}=\omega_{\ua\ua(\da\da)}$.
The two spin-flip excitations follow immediately as
\begin{eqnarray} \label{omegasf1}
E^{sf+} &=& \omega_{\ua\da}+  M_{\ua \da,\ua \da}
\\
E^{sf-} &=& \omega_{\da\ua}+ M_{\da \ua,\da \ua} \:, \label{omegasf2}
\end{eqnarray}
where $E^{sf+}_{KS}=\omega_{\ua\da}$ and  $E^{sf-}_{KS}=\omega_{\da\ua}$.

This gives a simple approximation for the $xc$ correction to the spin
stiffness:
\begin{equation} \label{deltatddft}
\Delta_{xc}^s =  M_{\ua \da,\ua \da} + M_{\da \ua,\da \ua} \:.
\end{equation}
Explicit expressions for $f^{xc}_{\alpha \alpha',\sigma \sigma'}$ can be obtained from the local spin-density approximation (LSDA),
and we list them here for completeness (see also Wang and Ziegler \cite{Ziegler}):
\begin{eqnarray}
f^{xc}_{\ua \ua,\ua \ua} &=&  \frac{\partial^2(n e_{\rm xc}^h)}{\partial n^2}
+ 2(1-\zeta) \frac{\partial^2 e_{\rm xc}^h}{\partial n \partial \zeta}
+ \frac{(1-\zeta)^2}{n} \frac{\partial ^2 e_{\rm xc}^h}{\partial \zeta^2}
\nonumber \\
f^{xc}_{\da \da,\da \da} &=&  \frac{\partial^2(n e_{\rm xc}^h)}{\partial n^2}
- 2(1+\zeta) \frac{\partial^2 e_{\rm xc}^h}{\partial n \partial \zeta}
+ \frac{(1+\zeta)^2}{n} \frac{\partial ^2 e_{\rm xc}^h}{\partial \zeta^2}
\nonumber \\
f^{xc}_{\ua \ua,\da \da} &=&  \frac{\partial^2(n e_{\rm xc}^h)}{\partial n^2}
- 2\zeta \frac{\partial^2 e_{\rm xc}^h}{\partial n \partial \zeta}
- \frac{(1-\zeta^2)}{n} \frac{\partial ^2 e_{\rm xc}^h}{\partial \zeta^2}
\nonumber \\
f^{xc}_{\ua \da,\ua \da} &=&
\frac{2}{n\zeta} \frac{\partial e_{\rm xc}^{h}(n,\zeta)}{\partial \zeta} \:,
\end{eqnarray}
where $f^{xc}_{\da \da,\ua \ua} = f^{xc}_{\ua \ua,\da \da}$ and $f^{xc}_{\da \ua,\da \ua}=f^{xc}_{\ua \da,\ua \da}$,
and it is understood that all expressions are multiplied by $\delta (\bfr - \bfr')$ and evaluated at
the local ground-state density and spin polarization,
$n_0(\bfr) = n_{0\ua}(\bfr) + n_{0\da}(\bfr)$  and $\zeta_0(\bfr) = [n_{0\ua}(\bfr) - n_{0\da}(\bfr)]/n_0(\bfr)$.
For the $xc$ energy density of the spin-polarized homogeneous
electron gas we take the standard interpolation formula
\begin{eqnarray}
e_{xc}^h(n,\zeta) &=& e_{xc}^h(n,0) + \frac{(1+\zeta)^{4/3} + (1-\zeta)^{4/3} - 2}{2^{4/3}-2}
\nonumber\\
&\times& [e_{xc}^h(n,1)-e_{xc}^h(n,0)].
\end{eqnarray}
The case of exact exchange (XX) in linear response can be treated exactly, though with considerable technical and
numerical effort.\cite{Kim2002,Hellgren2008} A simplified expression of the XX {\em xc} kernel
was developed by Petersilka {\em et al.}\cite{Petersilka1996}, and we have generalized their expression
for the linear response of the spin-density matrix. We obtain (details will be published elsewhere):
\begin{equation}
f^x_{\ua\ua,\ua\ua}(\bfr,\bfr') =
-\sum_{i,k}^{N_\ua}
\frac{\psi_{k\ua}(\bfr)\psi_{k\ua}^*(\bfr') \psi^*_{i\ua}(\bfr) \psi_{i\ua}(\bfr')}
{|\bfr - \bfr'|n_{\ua}(\bfr) n_{\ua}(\bfr')}
\end{equation}
and similarly for $f^x_{\da\da,\da\da}(\bfr,\bfr')$, and
\begin{eqnarray}
f^x_{\ua\da,\ua\da}(\bfr,\bfr') &=&
-\sum_{i,k}^{N_\ua,N_\da}
\frac{\psi_{k\ua}(\bfr)\psi_{k\ua}^*(\bfr') \psi^*_{i\da}(\bfr) \psi_{i\da}(\bfr')}
{|\bfr - \bfr'|\sqrt{n_{\ua}(\bfr)n_\da(\bfr) n_{\ua}(\bfr')n_\da(\bfr')}} \nonumber\\
&=&f^x_{\da\ua,\da\ua}(\bfr,\bfr').
\end{eqnarray}
Here, $N_\ua$ and $N_\da$ are the number of occupied spin-up and spin-down orbitals.

\begin{table}
\caption{\label{table3} Top part: Spin-conserving and spin-flip excitation energies, calculated with
LSDA and KLI-XX using differences of KS eigenvalues and TDDFT in the single-pole approximation
(\ref{omegasc})--(\ref{omegasf2}). Bottom part: TDDFT {\em xc} corrections to the KS spin-flip excitation energies,
from Eqs. (\ref{omegasf1}), (\ref{omegasf2}),
and to the KS spin gap, Eq. (\ref{deltatddft}).
All numbers are in eV.}
\begin{ruledtabular}
\begin{tabular}{lcccccc}
 & \multicolumn{2}{c}{LSDA} & \multicolumn{2}{c}{KLI-XX} & \multicolumn{2}{c}{Exact}\\
 & KS & TDDFT & KS & TDDFT  & KS\footnote{Evaluated from the KS eigenvalues of
Ref.~\onlinecite{krieger}} & Exp\footnote{Spectroscopic data from
Ref.~\onlinecite{nist} ($E^{sc\ua,\da}$)
and Refs.~\onlinecite{Mannervik1983,Hsu1991} ($E^{sf+}$)
}\\ \hline
$E^{sc\ua}$ & $1.83$ & $2.00$ & $1.84$ & $2.01$ & $1.85$ & $1.85$\\
$E^{sc\da}$ & $48.72$  & $48.89$ & $58.90$ & $59.31$ & $56.25$ & $56.36$ \\
$E^{sf+}$ & $49.47$  & $48.23$ & $63.64$ & $62.12 $ & $60.87$ & $57.41$ \\
$E^{sf-}$ & $1.07$ & $0.99$ & $-2.89$ & $-2.97$ & $-2.77$ & $0.0$ \\
\\
$\Delta_{xc}^{sf+}$   & \multicolumn{2}{c}{$-1.24$} & \multicolumn{2}{c}{$-1.52$} & \multicolumn{2}{c}{$-3.46$}\\
$\Delta_{xc}^{sf-}$   & \multicolumn{2}{c}{$-0.08$} & \multicolumn{2}{c}{$-0.07$} & \multicolumn{2}{c}{$+2.77$}\\
$\Delta_{xc}^s    $   & \multicolumn{2}{c}{$-1.32$} & \multicolumn{2}{c}{$-1.59$} & \multicolumn{2}{c}{$-0.69$}
\end{tabular}
\end{ruledtabular}
\end{table}

We have evaluated Eqs. (\ref{omegasc})--(\ref{deltatddft}) for the spin-con\-serving and spin-flip
excitation energies of the Lithium atom involving the $H (\sigma)$ and $l(\sigma')$ orbitals.
The LSDA and KLI-XX orbital eigenvalues that are needed as input are given in Table \ref{table1}.

The associated excitation energies are shown in Table \ref{table3}, where we compare KS excitations,
i.e., differences of KS eigenvalues, with TDDFT excitations obtained using the single-pole approximation
described above. All in all, the TDDFT excitation energies are not much improved compared to the KS
orbital eigenvalue differences. The main reason is that the LSDA and KLI-XX KS energy eigenvalues are not
particularly close to the exact KS energy eigenvalues, and furthermore that the single-pole approximation
is too simplistic for this open-shell atom.

However, we observe that the $xc$ correction $\Delta_{xc}^s$ to the spin
stiffness $E_s$, when directly calculated within LSDA or KLI-XX using the TDDFT formula (\ref{deltatddft}), is reasonably
close to the exact value, and has the correct sign. This tells us that, even though the KS spin gap itself may be not very good,
the simple TDDFT expression (\ref{deltatddft}) gives a reasonable approximation for the $xc$ correction to it.


\section{Conclusion}
\label{concl}

The calculation of spin gaps and related quantities is important for
phenomena like spin-flip excitations in finite systems,\cite{Ziegler}
the magnetic and transport properties of extended systems such as
half-metallic ferromagnets\cite{coey} and, quite generally, in the
emerging field of spintronics and spin-dependent transport.

Our aim in this paper was to show how to define and calculate spin gaps and
related quantities from density-functional theory. The proper definition
of spin gaps in SDFT is by no means obvious, and the straightforward
extrapolation of concepts and properties from the charge case to the spin
case is fraught with dangers. Therefore, we started our investigation by
disentangling two aspects of the gap problem that in the charge case are
usually treated together: the derivative discontinuity and the many-body
correction to single-particle gaps.

On this background, we then provided a set of DFT-based definitions of
quantities that are related to spin gaps, such as spin-conserving gaps,
spin-flip gaps and the spin stiffness, pointing out in each case where
possible analogies to the charge case exist, and when these analogies break
down. In particular, spin-flips involve excitations, while particle addition
and removal involves ground-state energies. As a consequence,
single-particle spin-flip energies involve two eigenvalues (and not one)
and single-particle spin gaps involve four (and not two). Moreover,
each spin-flip energy may have its own $xc$ correction (there is no
Koopmans' theorem for spin flips).

An evaluation of our definitions for the Lithium atom, making use of
highly precise Kohn-Sham eigenvalues and spectroscopic data, shows that the
many-body correction to spin gaps can indeed be nonzero. In fact, unlike
what is common in the charge case, this correction turns out to be negative,
{\em i.e.} the single-particle calculation overestimates the spin gap while
it underestimates the charge gap. While this result for a single atom is
consistent with available data on half-metallic ferromagnets,\cite{coey}
similar calculations must be performed for other systems before broad trends
can be identified.

Next, we connected the many-body corrections to the spin gap and related
quantities to ensemble DFT and to TDDFT. The former connection
makes use of a suitable excited-state spin ensemble (different from the
degenerate-state spin ensemble recently proposed by Yang and collaborators
\cite{yang1,yang2}) and depends on a crucial insight of Levy \cite{levy}
regarding excited-state derivative discontinuities.
The latter connection employs a noncollinear version of the Casida
equations,\cite{Ziegler} which we evaluate, again for the Lithium atom,
within the single-pole approximation, in LSDA and for exact exchange.

The development of approximate density functionals and computational
methodologies that permit the reliable calculation of spin gaps and
related quantities, including their many-body ($xc$) corrections, remains
a challenge for the future.

{\bf Acknowledgments} K.C. thanks the Physics Department of the University
of Missouri-Columbia, where part of this work was done, for generous
hospitality, and D. Vieira for preparing and discussing Fig.~\ref{fig1}.
K.C. is supported by Brazilian funding agencies FAPESP and CNPq.
C.A.U. acknowledges support from NSF Grant No. DMR-0553485.
C.A.U. would also like to thank the KITP Santa Barbara for its hospitality
and partial support under NSF Grant No. PHY05-51164.
G.V. acknowledges support from NSF Grant No. DMR-0705460.

\end{document}